\documentclass[preprint,aps,amsmath,nofootinbib,tightenlines]{revtex4}
\usepackage{graphicx}
\usepackage{bm}
\usepackage{epsfig}
\usepackage{graphics}
\usepackage{amsfonts,amssymb}

\def\Dsl{\hbox{/\kern-.6000em D}} 

\def\dsl{\,\raise.15ex\hbox{/}\mkern-13.5mu D}
\def\bsigma{\mbox{\boldmath $\sigma$}}

\def\bsigma{\mbox{\boldmath $\sigma$}}

\def\ltap{\ \raise.3ex\hbox{$<$\kern-.75em\lower1ex\hbox{$\sim$}}\ }
\def\gtap{\ \raise.3ex\hbox{$>$\kern-.75em\lower1ex\hbox{$\sim$}}\ }
\def\OMIT#1{}

\def\lsim{\mathrel{\raise.3ex\hbox{$<$\kern-.75em\lower1ex\hbox{$\sim$}}}}
\def\gsim{\mathrel{\raise.3ex\hbox{$>$\kern-.75em\lower1ex\hbox{$\sim$}}}}

\def\msb{{\overline{\rm MS}}}

\newcommand{\nn}{\nonumber}

\newcommand{\bmk}{\mathbf k}
\newcommand{\bmp}{\mathbf p}

\newcommand{\bmA}{\mathbf A}

\newcommand{\bmnabla}{\mathbf \nabla}

\newcommand{\bmpp}{{\bmp^\prime}}

\newcommand{\as}{\alpha_S}
\newcommand{\au}{\alpha_U}

\newcommand{\dn}{\nu \frac{d}{d \nu}}
\def\msb{{\overline{\rm MS}}}

\catcode`\@=11
\def\slash{\mathpalette\make@slash}
\def\make@slash#1#2{\setbox\z@\hbox{$#1#2$}%
  \hbox to 0pt{\hss$#1/$\hss\kern-\wd0}\box0}
\catcode`\@=12 

\begin{document}


\preprint{ \vbox{ \hbox{MPP-2006-160} 
}}

\title{\phantom{x}\vspace{0.5cm} 
Two-Loop Ultrasoft Running of the $O(v^2)$ QCD Quark Potentials
\vspace{1.0cm} }

\author{Andr\'e~H.~Hoang and Maximilian~Stahlhofen \vspace{0.5cm}}
\affiliation{Max-Planck-Institut f\"ur Physik\\
(Werner-Heisenberg-Institut), \\
F\"ohringer Ring 6,\\
80805 M\"unchen, Germany\vspace{1cm}
\footnote{Electronic address: ahoang@mppmu.mpg.de, stahlhof@mppmu.mpg.de}\vspace{1cm}}


\begin{abstract}
\vspace{0.5cm}
\setlength\baselineskip{18pt}
The two-loop ultrasoft contributions to the next-to-leading logarithmic (NLL)
running of the QCD potentials at order $v^2$ are determined. The results
represent an important step towards the next-to-next-to-leading
logarithmic (NNLL) description of heavy quark pair production and annihilation close
to threshold.  
\end{abstract}
\maketitle


\newpage

\section{Introduction}
\label{sectionintroduction}

The measurement of the line-shape of the total top-antitop quark
production cross section in the threshold  
region $\sqrt{s}\approx 2m_t$ is one of the major tasks within the top quark
physics program at the ILC. The most prominent quantity to be measured is the
top quark mass, and one expects an improvement in precision of $m_t$ by 
about an order of magnitude to mass measurements based on reconstruction
obtained at the Tevatron and the LHC~\cite{thresholdscan}.

To obtain a meaningful theoretical description of the nonrelativistic
threshold dynamics it is required to systematically sum the so-called Coulomb
singular terms $\propto (\alpha_s/v)^n$ in a systematic
nonrelativistic expansion, $v$ being the relative velocity of the top quarks. This 
task is achieved by means of effective theories based on nonrelativistic QCD
(NRQCD)~\cite{BBL}. In this approach, however, sizeable logarithmic terms
$\propto(\alpha_s\ln v)^n$ are not systematically accounted for, which leads
to rather large normalization uncertainties of the cross section
line-shape. The next-to-next-to-leading order predictions in this
``fixed-order'' approach were estimated to have a normalization uncertainty of
order $20\%$~\cite{synopsis}. These normalization uncertainties do not affect the top
mass measurement, which primarily depends on the c.m. energy where the cross
section rises, but they render measurements of other quantities such as the
top total width or the top quark couplings impossible. To match the 
statistical uncertainties, that are expected for these quantities, a
theoretical precision of the cross section normalization of at least $3\%$ 
would be required.

In Refs.~\cite{HMST} (see also Ref.~\cite{PinedaSigner}) it was
demonstrated that the 
summation of logarithmic $(\alpha_s\ln v)^n$ terms, using
renormalization-group-improved perturbation theory, can significantly reduce the
normalization uncertainties of the threshold cross section. 
Concerning QCD effects the renormalization group improved LL
(leading-logarithmic) and NLL order predictions of the threshold cross section
are completely known, but no full NNLL order prediction exists at present. 
The full NNLL order prediction is, however, required to obtain a
reliable estimate of the remaining theoretical normalization uncertainties. 

The missing ingredient is the NNLL running of the Wilson coefficient of the
leading order effective current, that describes production and annihilation of
a nonrelativistic $t\bar t$ pair in a S-wave spin-triplet state
(${}^3S_1$). Adopting the label notation from
vNRQCD~\cite{LMR,HoangStewartultra} the current has the form
\begin{eqnarray} 
  {\bf J}_{1,\bf p} = 
    \psi_{\bmp}^\dagger\, \bsigma (i\sigma_2) \chi_{-\bmp}^*
\,,
\label{J1J0}
\end{eqnarray}
where  $\psi_{\bmp}$ and $\chi_{\bmp}$ annihilate top and antitop quarks with
three-momentum $\bmp$, respectively, and where color indices have been
suppressed. The current does not have a LL anomalous dimension because there
is no one-loop vertex diagram in the effective theory that contains UV
divergences associated with the current ${\bf J}_{1,\bf p}$. Such UV
divergences arise at NLL order from insertions of the NNLL kinetic energy
operators and from insertions of the NNLL order potentials~\cite{LMR}. The
corresponding computations were carried out in
Refs.~\cite{amis,amis2,Pineda:2001et,HoangStewartultra} and are 
completed. Using the conventions from~\cite{HoangStewartultra} the
resulting NLL order renormalization group equation for the Wilson
coefficient $c_1$ of the current has the form (${\bf S}^2=2$)
\begin{eqnarray}
 \nu \frac{\partial}{\partial\nu} \ln[c_1(\nu)] & = &
 -\:\frac{{\cal V}_c^{(s)}(\nu)
  }{ 16\pi^2} \bigg[ \frac{ {\cal V}_c^{(s)}(\nu) }{4 }
  +{\cal V}_2^{(s)}(\nu)+{\cal V}_r^{(s)}(\nu)
   + {\bf S}^2\: {\cal V}_s^{(s)}(\nu)  \bigg] 
   \nn\\
  && +\: \alpha_s^2(m\nu)\,\bigg[ \frac{C_F}{2}(C_F-2\,C_A)\bigg] 
     +  \alpha_s^2(m\nu)\,\bigg[  
     3 {\cal V}_{k1}^{(s)}(\nu) + 2 {\cal V}_{k2}^{(s)}(\nu) \bigg]
\,,
\label{c1anomdim}
\end{eqnarray}
where $\nu$ is the vNRQCD velocity renormalization parameter that is
conveniently used to parametrize the correlation between soft and ultrasoft
dynamical scales within the renormalized effective
theory~\cite{LMR}. The analogous
evolution equation  for currents describing pairs of quarks and colored scalars 
in any angular momentum and spin state (${}^{2s+1}L_J$) were
derived recently in Ref.~\cite{HoangRuiz2}. In
Eq.~(\ref{c1anomdim}) the term ${\cal V}_c^{(s)}$ is the Wilson
coefficient of the 
Coulomb potential $\propto 1/\bmk^2$, and ${\cal V}_2^{(s)}$ and ${\cal
  V}_r^{(s)}$ are the coefficients of the ${\cal O}(v^2)$ potentials with the
momentum structure $1/m^2$ and $(\bmp^2+{\bmp^{\prime}}^2)/(2m^2\bmk^2)$,
respectively, $m$ being the heavy quark mass. The coefficients ${\cal
  V}_{k1,k2}^{(s)}$ are from the so-called 
non-Abelian potentials that scale like $1/(m|\bmk|)$, and ${\cal V}_s^{(s)}$
is the coefficient of the spin-dependent potential that can contribute for
spin triplet S-wave states. The superscripts ${(s)}$ refer to the color singlet state of
the quark pair. 

At NNLL order there are two kinds of contributions to the evolution of $c_1$ that have to be accounted
for. The first kind arises from three-loop vertex diagrams that come from
insertions of subleading soft matrix element corrections to the potentials and
from insertions of potentials with additional exchange of ultrasoft
gluons. The corresponding computations were carried out in Ref.~\cite{3loop}. 
These effects are referred to as the non-mixing contributions as they 
affect the evolution of $c_1$ directly. The
second kind of contributions arises from the subleading evolution of the
potential Wilson coefficients that appear in the NLL order renormalization
group equation shown in Eq.~(\ref{c1anomdim}). They are
referred to as the mixing contributions as they affect the evolution of $c_1$ indirectly. 
Except for the coefficient of the Coulomb
potential ${\cal V}_c^{(s)}$~\cite{Pineda:2001ra,HoangStewartultra} and for
the spin-dependent potential ${\cal V}_s^{(s)}$~\cite{Penin:2004xi} no
complete determination for the subleading evolution 
exists at present. The analysis of the three-loop
(non-mixing) terms in Ref.~\cite{3loop} revealed that the contributions
involving the exchange of ultrasoft gluons are more than an order of magnitude
larger than those arising from soft matrix element insertions and are as large
as the known NLL contributions. The reason is
related to the larger size of the ultrasoft coupling $\alpha_s(m \nu^2)$ and to a rather large
coefficient multiplying the ultrasoft contributions. These large ultrasoft
contributions are responsible for an uncertainty in the normalization of the
most up-to-date threshold cross section prediction of at best
$6\%$~\cite{HoangEpiphany}, which is quite far from the required precision.
(See also Ref.~\cite{PinedaSigner} for an alternative analysis without NNLL
non-mixing effects.) The 
corresponding effects of the soft matrix element corrections are only at the 
level of several per mille. From this analysis it is reasonable to assume
that the ultrasoft effects, which form a gauge-invariant subset, also dominate
the mixing contributions. This assumption is also consistent with the
results for the NLL evolution of the 
coefficient ${\cal V}_s^{(s)}$~\cite{Penin:2004xi}, which is dominated by soft
effects, since the ultrasoft gluon coupling to heavy quarks is spin-independent
at this 
order, and which was found to have a very small numerical effect
as well~\cite{Penin:2004ay}. From the parametric point of view this can
be understood from the fact that ultrasoft effects can affect the
coefficients of the spin-dependent potentials only indirectly 
through mixing via potential loop divergences, since at this order the
coupling of ultrasoft gluons to heavy quarks are spin-independent.
For the coefficient of spin-dependent potentials this leads to
ultrasoft NLL terms $\propto\alpha_S^2 (\alpha_U\ln v)^n$, where
$\alpha_S$ and $\alpha_U$ are the strong coupling at the soft and
ultrasoft scales, respectively. These terms are parametrically suppressed 
compared to ultrasoft NLL terms $\propto\alpha_S\alpha_U (\alpha_U\ln
v)^n$ that can affect the spin-independent potential coefficients
through two-loop ultrasoft loop diagrams. 
We can therefore expect that ultrasoft effects are dominating
the higher order evolution of the spin-independent potentials.

In this work we determine as a first step the two-loop ultrasoft contributions
to the NLL anomalous dimensions of the spin-independent potentials
${\cal V}_2$ and ${\cal V}_r$. The outline of this work is as follows:
in Sec.~\ref{sectiontheory} we briefly review the effective theory
setup used for our work and in Sec.~\ref{sectionresults} we present
our results. Sec.~\ref{sectiondiscussion} contains a brief
numerical analysis and Sec.~\ref{sectionconclusion} the conclusion.

\section{Theoretical Setup}
\label{sectiontheory}

The vNRQCD Lagrangian basically consists of three parts \cite{LMR, HoangStewartultra},
\begin{equation}
 {\cal L}_{\rm vNRQCD}\:=\:  {\cal L}_{u} \:+\:  {\cal L}_{p} \:+\:  {\cal
   L}_{s}\,,
\end{equation}
containing kinetic terms and ultrasoft interactions, potential interactions
and interactions involving soft degrees of freedom, respectively. The
``ultrasoft part'' ${\cal L}_u$  has the form
\begin{equation}
{\cal L}_u  = 
\sum_{\mathbf p} \bigg\{
   \psi_{\bmp}^\dagger   \!\bigg[ i D^0 - \frac {\left({\bf p}\!-\!i{\bf D}\right)^2}
   {2 m}
 + \frac{\bmp^4}{8 m^3}
 + \ldots \!\bigg]\! \psi_{\bmp}
 + (\psi \!\to\! \chi,\, T \!\to\! \bar T) \bigg\}\!
 -\frac{1}{4}G^{\mu\nu}G_{\mu \nu} 
+\ldots ,\!
\label{Lus}
\end{equation}
where the ultrasoft gauge-covariant derivative reads 
$D^\mu = \partial^\mu + i g_U A^\mu(x)$ and $g_U$ is the strong coupling at
the ultrasoft scale $m\nu^2$. The terms describe besides the propagation of 
the heavy quarks their interaction with ultrasoft gluons. 
The ``potential part'' ${\cal L}_{p}$ describes
the  potential quark-antiquark interactions and reads 
\begin{eqnarray}
{\cal L}_p&=& - \sum_{\bmp, \bmpp}  
V_{\alpha\beta\lambda\tau} \left({\bmp},{\bmpp}\right)\ 
\psi_{\bmpp \alpha}^\dagger\: \psi_{\bmp\, \beta}\:
  \chi_{-\bmpp \lambda}^\dagger\:  \chi_{-\bmp\, \tau} + \ldots \,, 
\label{Lpot}
\end{eqnarray}
where $\alpha,\beta,\lambda,\tau$ are color indices and 
\begin{eqnarray}
V_{\alpha\beta\lambda\tau}({\bmp},{\bmp^\prime}) &=& 
(T^A_{\alpha\beta} \otimes \bar T^A_{\lambda\tau})\, \bigg[
 \frac{{\cal V}_c^{(T)}}{\bmk^2}
 + \frac{{\cal V}_k^{(T)}\pi^2}{m|{\bmk}|}
 + \frac{{\cal V}_r^{(T)}({\bmp^2 + \bmp^{\prime 2}})}{2 m^2 \bmk^2}
 + \frac{{\cal V}_2^{(T)}}{m^2}
 + \ldots \bigg] \nn \\
&& + \,(1_{\alpha\beta}\otimes \bar 1_{\lambda\tau})\, \bigg[
 \frac{{\cal V}_c^{(1)}}{\bmk^2}
 + \frac{{\cal V}_k^{(1)}\pi^2}{m|{\bmk}|}
 + \frac{{\cal V}_r^{(1)}({\bmp^2 + \bmp^{\prime 2}})}{2 m^2 \bmk^2}
 + \frac{{\cal V}_2^{(1)}}{m^2}
 +\ldots \bigg]
\label{pots}
\end{eqnarray}
with $\bmk = \bmpp-\bmp$.
The spin-dependent ${\cal O}(v^2)$ potentials have not been written
out in Eq.~(\ref{pots}), since they will not be relevant in this work.
One can convert the potentials from the basis formed by the two ${\bf
  3}\otimes{\bf \bar 3}$ color structures $1\otimes \bar 1$ and $T^A \otimes 
\bar T^A$ into the more physical color singlet/octet basis by 
\begin{eqnarray}
 \left[\begin{array}{c} V_{\rm singlet} \cr V_{\rm octet} \end{array}\right]
 =\left[\begin{array}{ccc} 1 &  & -C_F \cr
    1 &  & \frac{1}{2} C_A - C_F \cr
 \end{array}\right]
 \left[\begin{array}{c} V_{1\otimes 1} \cr V_{T\otimes T} 
 \end{array}\right]\,.
\label{convertpotentials}
\end{eqnarray}

\begin{figure}[ht]
\begin{center}
a)
\includegraphics[width = 0.15 \textwidth]{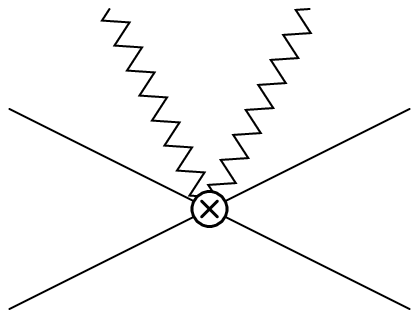} \hspace{0.1 \textwidth}
b)
\includegraphics[width = 0.15 \textwidth]{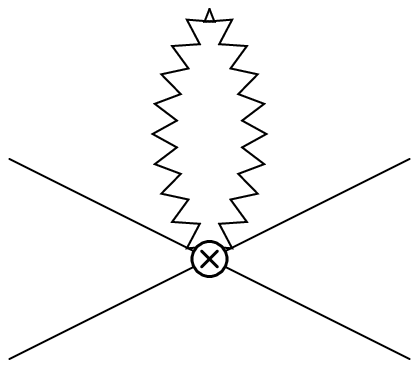} \hspace{0.1 \textwidth}
c)
\includegraphics[width = 0.15 \textwidth]{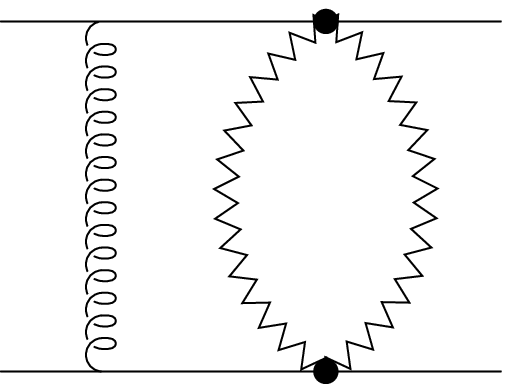}
\caption{a) Operator ${\cal O}_{2i}$, b) operator ${\cal O}_{2i}$ with soft lines 
closed to a loop, c) diagram with soft and ultrasoft loop contributing to potential 
counterterms. \label{softdiags} }
\end{center}
\end{figure}
In the soft sector of the theory there are 6-field operators needed for the
description of soft Compton scattering off quark-antiquark potentials (see
Fig.~\ref{softdiags}a).
As was pointed out in Ref.~\cite{HoangStewartultra} these operators have zero
matching conditions at the hard scale $\nu=1$ and contribute to the ultrasoft
renormalization of the ${\cal O}(v^2)$ potentials. The operators have the structure
\begin{eqnarray} 
 {\cal O}_{2\varphi}^{(\sigma),(T)} &=& { g_S^4}\:  \:
   (\psi_{\bmp^\prime}^\dagger\: \Gamma^{(\sigma),(T)}_{\varphi,\psi}\: 
   \psi_{\bmp}) \:
   (\chi_{-\bmp^\prime}^\dagger\: \Gamma^{(\sigma),(T)}_{\varphi,\chi}\: 
   \chi_{-\bmp})
   \ (\overline\varphi_{-q}\: \Gamma_\varphi^{(\sigma),(T)}\: \varphi_{q}) \,, \nn\\
 {\cal O}_{2A}^{(\sigma),(T)}  &=& { g_S^4}\:  \:
   (\psi_{\bmp^\prime}^\dagger\: \Gamma^{(\sigma),(T)}_{A,\psi} \: \psi_{\bmp})  \:
   (\chi_{-\bmp^\prime}^\dagger\: \Gamma^{(\sigma),(T)}_{A,\chi} \: \chi_{-\bmp})
   \ (A^\mu_{-q}\: \Gamma_{A,\mu\nu}^{(\sigma),(T)}\: A^\nu_{q}) \,,\\\
 {\cal O}_{2c}^{(\sigma),(T)}  &=& { g_S^4}\:  \:
   (\psi_{\bmp^\prime}^\dagger\: \Gamma^{(\sigma),(T)}_{c,\psi} \: \psi_{\bmp})  \:
   (\chi_{-\bmp^\prime}^\dagger\: \Gamma^{(\sigma),(T)}_{c,\chi} \: \chi_{-\bmp})
   \ (\bar c_{-q}\: \Gamma_c^{(\sigma),(T)}\: c_{q})
\,,
\label{6q}
\end{eqnarray}
where $\varphi_{q}, A^\mu_{q}$ and $c_{q}$ denote soft quarks, gluons and
ghosts respectively, $g_S$ is the strong coupling at the soft scale
$m\nu$ and $\sigma$ is the order at which the operators contribute in the
$v$-counting. For the explicit form of the the $\Gamma$'s we refer to
Ref.~\cite{HoangStewartultra} (see also Ref.~\cite{HoangRuiz1}). The
superscript $(T)$ 
refers to the color structure $(T^A \otimes \bar T^A)$ that arises upon
closing the two soft lines (see Fig.~\ref{softdiags}b). There is another 
set of such operators, which are not displayed, having the superscript $(1)$, 
that refers to the color structure $(1 \otimes \bar 1)$ that arises upon closing 
the two soft lines. As was shown in Ref.~\cite{HoangStewartultra}, closing the two 
respective two soft lines for the leading $\sigma=0$ operators, one obtains a 
structure for the one-loop four-quark matrix elements that is identical to the 
one-loop soft-matrix element corrections to the Coulomb potential. Moreover, the 
Wilson coefficients of the $\sigma=2$ operators become nonzero for $\nu<1$ from
ultrasoft UV divergences (and the associated pull-up
terms~\cite{HoangStewartultra,zerobin}) in 
higher-order corrections to the vNRQCD matrix elements describing Compton
scattering off a quark-antiquark pair. The coefficients of these operators
are directly affected by ultrasoft gluon corrections to this
process. Through the UV divergences that arise upon closing the two
soft lines they mix into the potential coefficients 
${\cal V}_2^{(s)}$ and ${\cal V}_r^{(s)}$. Using the counting mentioned above 
the contributions from these mixing
contributions to the evolution of ${\cal V}_2^{(s)}$ and ${\cal V}_r^{(s)}$
are parametrically suppressed compared to the contributions from the 
two-loop ultrasoft diagrams that renormalize these potentials directly.
For the purpose of this work we need the $\sigma=2$ operators that have the
form
\begin{eqnarray} 
 {\cal O}_{2k}^{(2),(1)} &=&  \frac{\bmk^2}{m^2}\, \sum_i {\cal O}_{2i}^{(0),(1)}
   \,,\quad
 {\cal O}_{2k}^{(2),(T)} \,= \, \frac{\bmk^2}{m^2}\, \sum_i {\cal O}_{2i}^{(0),(T)}
   \,,\quad \nn\\
 {\cal O}_{2p}^{(2),(1)} &=&  \frac{(\bmp^2+\bmp^{\prime\,2})}{m^2}\, 
   \sum_i {\cal O}_{2i}^{(0),(1)} \,,\quad
 {\cal O}_{2p}^{(2),(T)} \,=\,  \frac{(\bmp^2+\bmp^{\prime\,2})}{m^2}\, 
   \sum_i {\cal O}_{2i}^{(0),(T)} \,,
\label{O2kp}
\end{eqnarray}
which have the Wilson coefficients $C^{(1,T)}_{2k}$ and $C^{(1,T)}_{2p}$,
respectively. We note that the mixing effects of these operators lead to
numerical effects for ${\cal V}_2^{(s)}$ and ${\cal V}_r^{(s)}$ that are
substantially smaller than those from the two-loop ultrasoft diagrams that
renormalize ${\cal V}_2^{(s)}$ and ${\cal V}_r^{(s)}$ directly. We
include them because their contribution leads to a rather
simple form for the two-loop ultrasoft corrections to  ${\cal
  V}_2^{(s)}$ and ${\cal  V}_r^{(s)}$ we determine in this work and also in order 
to illustrate the size of the contributions not considered in this work.

All fields, couplings and Wilson coefficients in the above Lagrangian are to be
understood as bare quantities. For the renormalized ones we chose the
usual conventions in $d=4-2\epsilon$ dimensions ($\mu_S=m\nu, \mu_U=m\nu^2$): 
\begin{equation}
\begin{array}{ll}
V =\mu_S^{2\epsilon}(V_R \!+\! \delta V)\,,& C_{2i} = C_{2i}^R \!+\! \delta C_{2i} \,,\\
g_U \,=\, \mu_U^\epsilon \, g_U^R\,,& g_S \,=\, \mu_S^\epsilon \, g_S^R\,,\\
\psi_\bmp = Z^{1/2}_{\psi , \bmp}\; \psi_\bmp^R\,,& Z_{\psi , \bmp} = 1 \!+\! \delta  Z_{\psi , \bmp}\,,\;\;(\psi \to \chi),\\
A^\mu = Z_A^{1/2} \: A_R^\mu\;,& Z_A=1 \!+\! \delta Z_A\,.
\end{array}
\end{equation}
For convenience, we will drop the index $R$ throughout this paper and only
deal with $\msb$ renormalized quantities in the following. 

\section{Results}
\label{sectionresults}

In general the counter term of the potentials induced by ultrasoft
renormalization takes the form 
\begin{equation}
 \delta \vec{V} = A\,\vec{V} + \vec{C}\,,
\label{generalVcounter}
\end{equation}
where $\delta \vec{V}$, $\vec{V}$ and $\vec{C}$ are 2-vectors and A is a $2
\times 2$ matrix in the $(1\otimes \bar 1$, $T^A \otimes \bar T^A)$ space,
\begin{equation}
 \vec{V}=\left[\begin{array}{c} V_{1\otimes 1}  \cr V_{T\otimes T} \end{array}\right] \,.
\label{renormpot}
\end{equation}
From Eq.~\eqref{generalVcounter} one can deduce the general form of the
renormalization group equations for the potentials:
\begin{eqnarray}
\dn\,\vec{V} &=& - 2 \epsilon \, \vec{V} - (1+A)^{-1}\big(\, 2 \epsilon\, 
\vec{C} \,+\,\dn \, \vec{C} \,+\,(\dn A)\,\vec{V}\,\big)
\,.
\label{VRGE} 
\end{eqnarray}
The full one-loop results for $A$ and $\vec C$ can be found in
Ref.~\cite{amis,HoangStewartultra}. For the ultrasoft contributions 
at the two-loop level we define different classes of (divergent) two-loop
diagrams in Feynman gauge contributing to the matrix $A$:\\[3ex] 
\tabcolsep1ex
\begin{tabular}{ccp{0.7 \textwidth}c}
class & topology& insertions/vertices (e.g. $\psi_{\bmp}^\dagger\, g A^0
\psi_{\bmp}$,\, $\psi_{\bmp}^\dagger \,g \frac{\bmp \cdot \bmA}{m}
\psi_{\bmp}$,\, $\psi_{\bmp}^\dagger \frac{\bmp \cdot \bmnabla}{m}
\psi_{\bmp}$,\, etc.)  & order\\[2 ex] 
1& Fig.~\ref{AllDiags}a & four $g A^0$ vertices & ${\cal O}(v^0)$ \\[1 ex]
2& Fig.~\ref{AllDiags}b & one gluon selfenergy insertion and two $g A^0$
vertices & ${\cal O}(v^0)$ \\[1 ex] 
3& Fig.~\ref{AllDiags}a & two $g A^0$ and two $g \frac{\bmp \cdot \bmA}{m}$
vertices & ${\cal O}(v^2)$ \\[1 ex] 
4& Fig.~\ref{AllDiags}b & one gluon selfenergy insertion and two $g
\frac{\bmp \cdot \bmA}{m}$ vertices & ${\cal O}(v^2)$ \\[1 ex] 
5& Fig.~\ref{AllDiags}a & two $g A^0$ vertices and two insertions of the
$\frac{\bmp \cdot \bmnabla}{m}$ operator on the internal quark lines &
${\cal O}(v^2)$\\[4 ex] 
6& Fig.~\ref{AllDiags}b & one gluon selfenergy insertion, two $g A^0$
vertices and two insertions of the $\frac{\bmp \cdot \bmnabla}{m}$ operator on
the internal quark lines & ${\cal O}(v^2)$\\[4 ex] 
7& Fig.~\ref{AllDiags}b & one gluon selfenergy insertion, one $g A^0$, one
$g \frac{\bmp \cdot \bmA}{m}$ vertex and one insertion of the $\frac{\bmp
  \cdot \bmnabla}{m}$ operator on the internal quark lines & ${\cal O}(v^2)$\\[4 ex] 
8& Fig.~\ref{AllDiags}c & one triple gluon interaction, one $g A^0$ and two
$g \frac{\bmp \cdot \bmA}{m}$ vertices & ${\cal O}(v^2)$ \\[1 ex] 
9& Fig.~\ref{AllDiags}c & one triple gluon interaction, two $g A^0$, one $g
\frac{\bmp \cdot \bmA}{m}$ vertex and one insertion of the $\frac{\bmp \cdot
  \bmnabla}{m}$ operator on the internal quark lines  & ${\cal O}(v^2)$ \\[6 ex] 
\end{tabular}
\begin{figure}[ht]
 \includegraphics[width = 1 \textwidth]{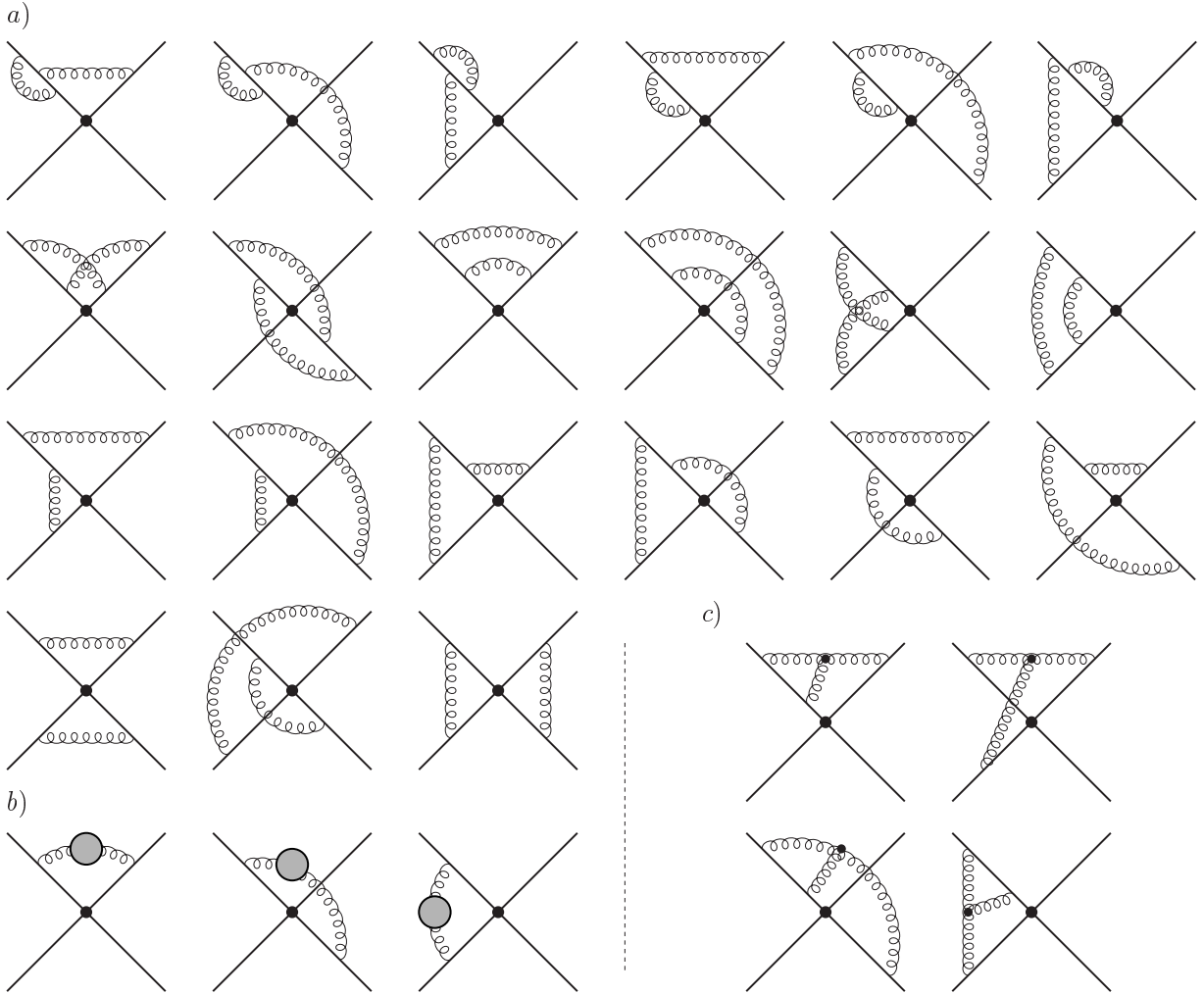}
\caption{2-loop topologies: a) ``Abelian'', b) with gluon selfenergy, c)
  ``Non-Abelian''. Left-right and up-down mirror graphs are not
  shown. \label{AllDiags}} 
\end{figure}
To each of these classes there is a corresponding class of diagrams, which
contribute to the heavy quark wavefunction counterterm $\delta Z_{\psi,\bmp}$,
with the same type and number of vertices and insertions. The results for the
different contributions to $\delta Z_{\psi,\bmp}$ after subtraction of the
respective one-loop subdivergencies are shown in Tab.~\ref{dZpsi}. The table
also shows an example diagram for each of the classes. 
\begin{table}[ht]
\begin{tabular}{|c|c|c|c|}
\hline
class & example diagram & contribution to $\delta Z_{\psi,\bmp}$ & order\\
\hline
\raisebox{2 ex}{1}& \begin{picture}(95,34)(0,3) \includegraphics[width = 100
    pt]{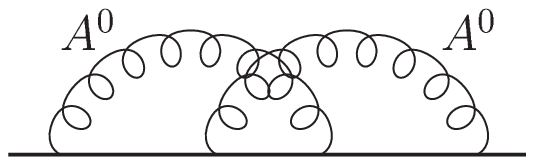} \end{picture} & \raisebox{2
  ex}{$\frac{C_F(C_F-C_A) \au^2}{8 \pi ^2 \epsilon ^2} + \frac{C_A C_F \au
    ^2}{8 \pi ^2 \epsilon }$} &  \raisebox{2 ex}{${\cal O}(v^0)$}\\ 
\hline
\raisebox{2 ex}{2}& \begin{picture}(95,34)(0,3) \includegraphics[width = 100
    pt]{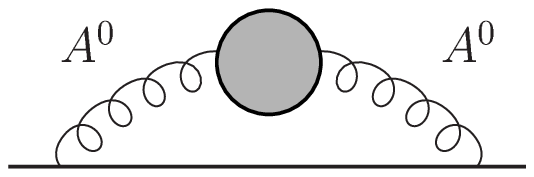} \end{picture} & \raisebox{2 ex}{$-\frac{C_F (5
    C_A - 4\,T n_f) \au^2}{32 \pi ^2 \epsilon ^2} + \frac{C_F (13 C_A-8\, T 
    n_f) \au^2}{48 \pi ^2 \epsilon } $} &  \raisebox{2 ex}{${\cal O}(v^0)$}\\ 
\hline
\raisebox{2 ex}{3}& \begin{picture}(95,34)(0,3) \includegraphics[width = 100
    pt]{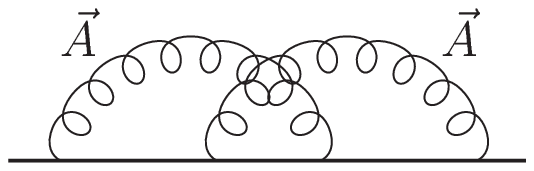} \end{picture} & \raisebox{2 ex}{$-\frac{C_F
    (C_F-C_A) \bmp^2 \au^2}{4\,m^2 \pi ^2 \epsilon ^2} - \frac{C_A C_F \bmp^2
    \au^2}{4\, m^2 \pi ^2 \epsilon } $} &  \raisebox{2 ex}{${\cal O}(v^2)$}\\
\hline
\raisebox{2 ex}{4}& \begin{picture}(95,34)(0,3) \includegraphics[width = 100
    pt]{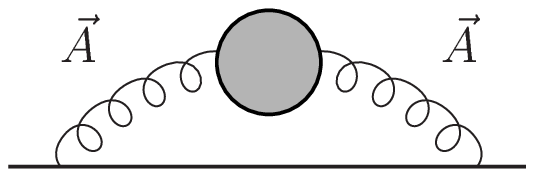} \end{picture} & \raisebox{2 ex}{$\frac{C_F
    (15\, C_A-12\, T n_f) \bmp^2 \au^2}{288\, m^2 \pi^2 \epsilon^2} -
  \frac{C_F (46\, C_A - 32\, T n_f)\bmp^2 \au ^2}{288\, m^2 \pi ^2 \epsilon }
  $} &  \raisebox{2 ex}{${\cal O}(v^2)$}\\ 
\hline
\raisebox{2 ex}{5}& \begin{picture}(95,34)(0,3) \includegraphics[width = 100
    pt]{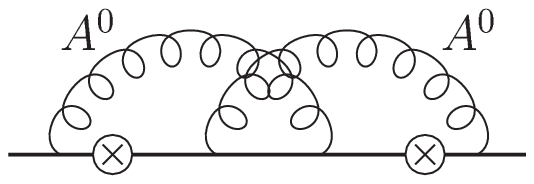} \end{picture} & \raisebox{2 ex}{$\frac{C_F
    (C_F-C_A) \bmp^2 \au^2}{4\,m^2 \pi ^2 \epsilon ^2} + \frac{C_A C_F \bmp^2
    \au^2}{4\, m^2 \pi ^2 \epsilon }$} & \raisebox{2 ex}{${\cal O}(v^2)$}\\ 
\hline
\raisebox{2 ex}{6}& \begin{picture}(95,34)(0,3) \includegraphics[width = 100
    pt]{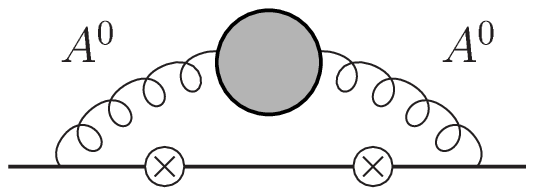} \end{picture} & \raisebox{2 ex}{$-\frac{C_F
    (75\,C_A-60\, T n_f) \bmp^2 \au^2}{288\, m^2 \pi ^2 \epsilon ^2} +
  \frac{C_F (110\,C_A-64\,T n_f) \bmp^2 \au^2}{288\, m^2 \pi^2 \epsilon }$} &
\raisebox{2 ex}{${\cal O}(v^2)$}\\ 
\hline
\raisebox{2 ex}{7}& \begin{picture}(95,34)(0,3) \includegraphics[width = 100
    pt]{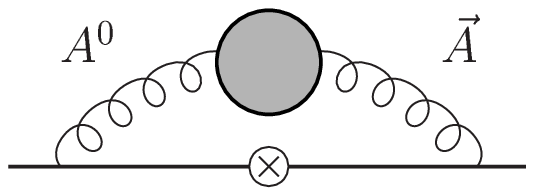} \end{picture} & \raisebox{2 ex}{$\frac{C_F
    (15\, C_A-12\, T n_f) \bmp^2 \au^2}{72\, m^2 \pi ^2 \epsilon ^2} -
  \frac{C_F (16\, C_A - 8\, T n_f) \bmp^2 \au^2}{72\, m^2 \pi ^2 \epsilon }$}
& \raisebox{2 ex}{${\cal O}(v^2)$}\\ 
\hline
\raisebox{2 ex}{8}& \begin{picture}(95,34)(0,3) \includegraphics[width = 100
    pt]{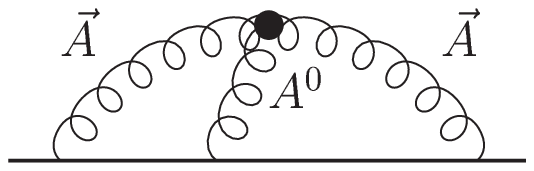} \end{picture} & \raisebox{2 ex}{0} &
\raisebox{2 ex}{${\cal O}(v^2)$} \\ 
\hline
\raisebox{2 ex}{9}& \begin{picture}(95,34)(0,3) \includegraphics[width = 100
    pt]{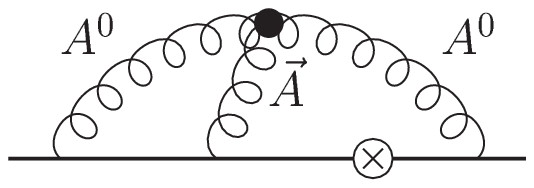} \end{picture} & \raisebox{2 ex}{0} &
\raisebox{2 ex}{${\cal O}(v^2)$}\\ 
\hline
\end{tabular}
\caption{Contributions to $\delta Z_{\psi,\bmp}$ at ultrasoft 2-loop level
  $(\au \equiv \alpha_s(m \nu^2))$. \label{dZpsi}} 
\end{table}
The result for the ultrasoft two-loop contributions to the matrix $A$ reads:
\begin{eqnarray}
A &=&
{\textstyle
\frac{\au^2}{m^2}
\left(
-\frac{11 C_A - 4 T n_f}{72\, \pi^2 \epsilon ^2}+\frac{(-3 \pi ^2 + 47) C_A -
  10 T n_f}{108\,\pi^2 \epsilon } 
\right)}
\left[
\begin{array}{ll}
 C_F \bmk^2 & - C_1 \bmk^2 \\
 -\bmk^2 & \left(C_F+\frac{C_d}{4}-\frac{3 C_A}{4}\right) \bmk^2+C_A
 (\bmp^2+\bmpp^2) 
\end{array}
\right] \nn\\[1 ex]
&+&
{\textstyle 
\frac{\au^2}{m^2}
\frac{C_A}{6 \epsilon }}
\left[
\begin{array}{ll}
 0 & C_1 \left(\bmk^2-\frac{9}{4} (\bmp^2+\bmpp^2)\right) \\
-\frac{1}{4} (\bmp^2+\bmpp^2)-\bmk^2 \;\; & -\frac{5}{8} (C_A - 4 C_F)
(\bmp^2+\bmpp^2) 
\end{array}
\right],
\label{Amatrix}
\end{eqnarray}
where $C_1 = \frac12 C_F (C_A - 2 C_F)$ and $C_d = 8 C_F - 3 C_A$.\\
Here and in the following $\au \equiv \alpha_s(m \nu^2)$ and $\as \equiv
\alpha_s(m \nu)$. The expression in Eq.~(\ref{Amatrix}) represents the
main result of this work. It is an important cross check that 
the ${\cal O}(\au^2\, v^0)$ terms vanish identically as a consequence of
ultrasoft gauge invariance. In addition, our result for the ultrasoft 
two-loop ${\cal O}(v^0)$ wave function renormalization constant agrees 
with the ${\cal O}(\alpha_s^2\, m^0)$ heavy quark wave function
renormalization constant in heavy quark effective theory (HQET), see
e.g. Ref.~\cite{Grozin:2000cm}. 
There is agreement since the purely ultrasoft sector of NRQCD is
closely related to HQET~\cite{Manohar:1997qy}. Furthermore we find that 
all ${\cal O}(v^2)$ terms cancel in our result for the ultrasoft
two-loop wave function renormalization constant (see
Tab.~\ref{dZpsi}). This cancellation also takes place for the  
${\cal O}(v^2)$ terms at one-loop and is required by consistency since
the form of the bilinear heavy quark kinetic terms in Eq.~(\ref{Lus}) 
is protected from quantum corrections by reparametrization invariance. 
As an additional cross check we have also determined all two-loop 
fermionic $Tn_f$ corrections in Coulomb gauge~\cite{StahlhofenTUM}. 

From the results obtained in Eq.~(\ref{Amatrix}) one can deduce
without effort the ultrasoft contributions to the NLL evolution of the
Wilson coefficients $C^{(1,T)}_{2k,2p}$, since the additional soft
external lines describing soft Compton scattering do not affect the
structure of the ultrasoft loop diagrams. 
Including the two-loop soft terms
induced by the pull-up mechanism in analogy to the 1-loop case
\cite{HoangStewartultra,zerobin} and the corresponding (known)
one-loop contributions the respective renormalization group
equations read
\begin{eqnarray}
\nu \frac{d}{d \nu} C_{2\bmk}^{(1,T)}(\nu) 
& = & 2\,A^{(1,T)}_{\bmk}\,(2\au-\as) + 
      4\,B^{(1,T)}_{\bmk}\,(2 \au^2 - \as^2)
\,,\\[2mm]
\nu \frac{d}{d \nu} C_{2\bmp}^{(1,T)}(\nu) 
& = & 2\,A^{(1,T)}_{\bmp}\,(2\au-\as) + 
      4\,B^{(1,T)}_{\bmp}\,(2 \au^2 - \as^2)
\,,
\end{eqnarray}
where the constants $A^{(1,T)}_{\bmk,\bmp}$ and $B^{(1,T)}_{\bmk,\bmp}$ are
defined as
\begin{eqnarray}
A^{(1)}_{\bmp} & = & 0 
\,,\qquad\qquad
B^{(1)}_{\bmp} \, = \, -\frac{3}{16}C_A C_F(C_A-2C_F)
\,,\nonumber\\[2mm]
A^{(T)}_{\bmp} & = & \frac{C_A}{3\pi}
\,,\qquad\quad
B^{(T)}_{\bmp} \, = \, C_A\frac{C_A(-57\pi^2+188)-40 Tn_f}{432\pi^2}+\frac{5
  C_A C_F}{12}
\,,\nonumber\\[2mm]
A^{(1)}_{\bmk} & = & -\frac{C_F(C_A-2C_F)}{6\pi}
\,,\qquad
B^{(1)}_{\bmk} \, = \, C_F(C_A-2C_F)\frac{C_A(21\pi^2-47)+10 Tn_f}{216\pi^2}
\,,\nonumber\\[2mm]
A^{(T)}_{\bmk} & = & -\frac{C_A-2C_F}{2\pi}
\,,\qquad\qquad
B^{(T)}_{\bmk} \, = \, (C_A-2C_F)\frac{C_A(3\pi^2-47)+10 Tn_f}{72\pi^2}
\,.
\end{eqnarray}
The solutions respecting the matching conditions
$C^{(1,T)}_{2k,2p}=0$ at $\nu=1$ have the form \newline ($\beta_0 = \frac{11}{3}C_A - \frac43 T
n_f \,,\; \beta_1 = \frac{34}{3}C_A^2 - 4 C_F T n_f - \frac{20}{3}C_A
T n_f$): 
\begin{eqnarray}
C_{2\bmp}^{(1,T)}(\nu) & = & 
-\frac{4\pi A^{(1,T)}_{\bmp}}{\beta_0} \ln\left(\frac{\au}{\as}\right)
+\left[\frac{A^{(1,T)}_{\bmp}\beta_1}{\beta_0^2}-
   \frac{8\pi B^{(1,T)}_{\bmp}}{\beta_0}
\right] (\au-\as) \,,
\nonumber\\[2mm]
C_{2\bmk}^{(1,T)}(\nu) & = & 
-\frac{4\pi A^{(1,T)}_{\bmk}}{\beta_0} \ln\left(\frac{\au}{\as}\right)
+\left[\frac{A^{(1,T)}_{\bmk}\beta_1}{\beta_0^2}-
   \frac{8\pi B^{(1,T)}_{\bmk}}{\beta_0}
\right] (\au-\as) \,.
\end{eqnarray}
Upon closing up the two soft gluon, ghost and light quark lines
one arrives at the following mixing contributions for 
$\vec C$ in Eq.~(\ref{generalVcounter}):
\begin{equation}
\vec{C} =  - \frac{\as^2 \,\beta_0}{\bmk^2 \epsilon} 
\left[\begin{array}{c} C_{2\bmk}^{(1)}\, \frac{\bmk^2}{m^2} \,+\,
    C_{2\bmp}^{(1)}\, \frac{\bmp^2+\bmpp^2}{m^2}  \\[1 ex]
    C_{2\bmk}^{(T)}\, \frac{\bmk^2}{m^2} \,+\, C_{2\bmp}^{(T)}\,
    \frac{\bmp^2+\bmpp^2}{m^2} \end{array}\right] .
\label{vecC}
\end{equation}
It is then straightforward to derive the evolution equation for the
coefficients ${\cal V}_2^{(s)}$ and ${\cal V}_r^{(s)}$ of the
spin-independent ${\cal O}(v^2)$ potentials. Using 
\begin{equation}
 \vec{V} =  \left[\begin{array}{c} 0 \\ \frac{{\cal V}_c(\nu)}{\bmk^2} \end{array}\right] 
\end{equation}
with ${\cal V}_c(\nu) \,=\, 4\, \pi\, \alpha_s(m\nu)$ on the RHS of
Eq.~(\ref{VRGE}), and including again the soft pull-up terms
we obtain 
\begin{eqnarray}
 \nu\,\frac{d}{d\nu}{\cal V}_2^{(1,T)}(\nu) & = &
8 \pi\, A_\bmk^{(1,T)}\,\as \,(2\au-\as) +
16\pi \, B_\bmk^{(1,T)}\,\as  \,(2 \au^2 - \as^2) -
2\,\beta_0\,\as^2\,C_{\bmk}^{(1,T)}
\,,\nonumber\\[2mm]
 \nu\,\frac{d}{d\nu}{\cal V}_r^{(1,T)}(\nu) & = &
8 \pi\, A_\bmp^{(1,T)}\,\as \,(2\au-\as) +
16\pi \, B_\bmp^{(1,T)}\,\as  \,(2 \au^2 - \as^2) -
2\,\beta_0\,\as^2\,C_{\bmp}^{(1,T)}
\,,\quad
\label{ddvV2r}
\end{eqnarray}
for the dominant NLL ultrasoft contributions.
To render the anomalous dimensions finite, we also have to account for UV-divergent 
contributions from diagrams of the type in Fig.~\ref{softdiags}c.
Note that in Eqs.~(\ref{ddvV2r}) we have
also displayed the LL ultrasoft terms including
the respective soft pull-up terms. The full set of the known LL
contributions from soft one-loop diagrams~\cite{amis} has not been
displayed.

From the previous equations we find the following simple result for the
ultrasoft one-loop and two-loop contributions to the evolution of the coefficients 
${\cal V}_2$ and ${\cal V}_r$ up to NLL order:
\begin{eqnarray}
\Big( {\cal V}_2^{(1,T)}(\nu) \Big)_{us} & = & 4\pi\,\alpha_s(m\nu)\,C_\bmk^{(1,T)}(\nu)
\,,\nonumber\\[2mm]
\Big( {\cal V}_r^{(1,T)}(\nu) \Big)_{us} & = & 8\pi\,\alpha_s(m\nu)\,C_\bmp^{(1,T)}(\nu)
\,.
\end{eqnarray}
The corresponding coefficients for the color singlet potentials read
\begin{eqnarray}
\Big( {\cal V}_2^{(s)}(\nu) \Big)_{us} & = & 4\pi\,\alpha_s(m\nu)\,
\Big(C_\bmk^{(1)}(\nu)-C_F C_\bmk^{(T)}(\nu)\Big)
\,,\nonumber\\[2mm]
\Big( {\cal V}_r^{(s)}(\nu) \Big)_{us} & = & 8\pi\,\alpha_s(m\nu)\,
\Big(C_\bmp^{(1)}(\nu)-C_F C_\bmp^{(T)}(\nu)\Big)\,.
\label{singres}
\end{eqnarray}
It is now straightforward to derive from Eq.~(\ref{c1anomdim}) 
the two-loop ultrasoft part of the NNLL mixing contributions to the
running of the ${}^3S_1$ current coefficient $c_1$ from the coefficients
${\cal V}_2^{(s)}$ and  ${\cal V}_r^{(s)}$. Parametrizing the form of
$c_1$ as~\cite{3loop}
\begin{eqnarray}
 \ln\Big[ \frac{c_1(\nu)}{c_1(1)} \Big] & = &
\xi^{\rm NLL}(\nu) + 
\Big(\,
\xi^{\rm NNLL}_{\rm m}(\nu) + \xi^{\rm NNLL}_{\rm nm}(\nu)
\,\Big) + \ldots
\,,
\label{c1solution}
\end{eqnarray}
where $\xi^{\rm NNLL}_m$ and $\xi^{\rm NNLL}_{nm}$ refer to the NNLL mixing
and non-mixing contributions, respectively, we find
\begin{eqnarray}
\xi^{\rm NNLL}_{m;2r,{\rm usoft}} & = &
\frac{2\pi C_F\beta_1}{\beta_0^3}\,\tilde A\,\alpha_s^2(m)\,
 \bigg[ -\frac{7}{4}+\frac{\pi^2}{6}+z\left(1-\ln\frac{z}{2-z}\right)
      +z^2\left(\frac{3}{4}-\frac{1}{2}\ln z\right)
\nonumber\\[2mm] & &\hspace{3cm} 
      -\ln^2\left(\frac{z}{2}\right)+\ln^2\left(\frac{z}{2-z}\right)
      -2\mbox{Li}_2\left(\frac{z}{2}\right)\bigg]
\nonumber\\[2mm] & &
+\,\frac{8\pi^2 C_F}{\beta_0^2}\,\tilde B\,\alpha_s^2(m)\,
  \Big[ 3-2z-z^2-4\ln(2-z) \Big]
\,,
\label{deltac1}
\end{eqnarray}
where
\begin{eqnarray}
\tilde A &= & A_\bmk^{(1)} + 2 A_\bmp^{(1)} 
  - C_F\Big( A_\bmk^{(T)} + 2 A_\bmp^{(T)}\Big)
\,,\nonumber\\[2mm]
\tilde B &= & B_\bmk^{(1)} + 2 B_\bmp^{(1)} 
  - C_F\Big( B_\bmk^{(T)} + 2 B_\bmp^{(T)}\Big)
\,,\nonumber\\[2mm]
z & \equiv & \left(\frac{\alpha_s(m\nu)}{\alpha_s(m)}\right)^{\rm LL}
  \, = \,
  \bigg(1+\frac{\alpha_s(m)\beta_0}{2\pi}\ln\nu\bigg)^{-1}
\,.
\end{eqnarray}
For completeness we also give the form of the ultrasoft corrections to
$\xi^{\rm NLL}$ from ${\cal V}_2$ and ${\cal V}_r$:
\begin{equation}
\xi^{\rm NLL}_{2r,{\rm usoft}} \:=\:
\frac{8\pi^2 C_F}{\beta_0^2}\,\tilde A\,\alpha_s(m)\,
 \Big[ -1+z+(2-z)\ln(2-z) \Big].
\end{equation}

\section{Numerical Discussion}
\label{sectiondiscussion}

\begin{figure}[ht]
\begin{center}
\epsfxsize=\textwidth
\epsffile[110 20 845 470]{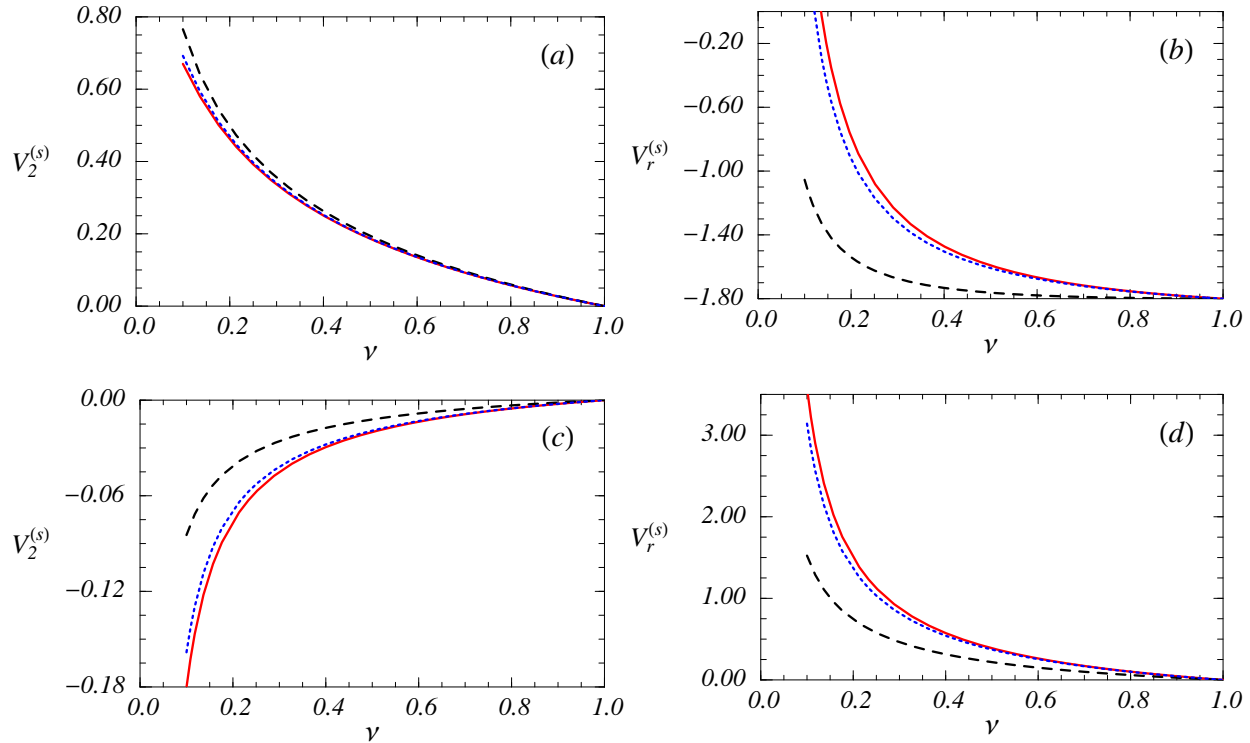}
\vspace{-1.3 cm}
\end{center}
\caption{Renormalization scale dependence of ${\cal V}_2^{(s)}$ and
 ${\cal V}_r^{(s)}$. In a), b) the full LL results (soft + ultrasoft) (dashed lines), 
 plus the NLL ultrasoft contributions with (solid lines) and without (dotted lines) 
the mixing terms is shown. In c), d) only the corresponding 
ultrasoft results are displayed omitting the known soft LL running. All of them include 
soft pull-up terms. \label{plots}} 
\end{figure}

In Fig.~\ref{plots}a,b the renormalization parameter $\nu$ dependence
of the color singlet coefficients
${\cal V}_2^{(s)}$ and ${\cal V}_r^{(s)}$ in Eq.~\eqref{singres} is
shown at LL order (dashed lines) including all soft and ultrasoft
contributions~\cite{HoangStewartultra,Pineda:2001ra}, and in addition
including the full two-loop 
ultrasoft contributions with (solid lines) and without the mixing contributions from
the zero matching operators 
(dotted lines), that are represented by the respective last terms on the 
RHS of Eqs.~\eqref{ddvV2r}. For the heavy quark
mass we have adopted $m_t=175$~GeV for the application to the top-antitop
quark threshold, and $\alpha_s(m_t)=0.1074$.
To further illustrate the importance of the two-loop ultrasoft
contributions we have displayed in Fig.~\ref{plots}c,d the evolution of the
coefficients at LL order due to the one-loop ultrasoft contributions
alone (dashed lines), and in addition including the two-loop
ultrasoft contributions analogous to Figs.~\ref{plots}a and b. 
Comparing the behavior of the LL order and NLL two-loop
ultrasoft contributions in Fig.~\ref{plots}c,d we find that the NLL
order corrections are of the same size as the ultrasoft LL
order contributions.
While for the potential ${\cal V}_2^{(s)}$ this
is of no concern, since for ${\cal V}_2^{(s)}$ the ultrasoft
contributions are suppressed by a small overall color factor compared to the
soft LL contributions, the result for ${\cal V}_r^{(s)}$ shows
that the two-loop ultrasoft corrections are indeed anomalously large, 
similar to the NNLL ultrasoft non-mixing contributions to the
evolution of $c_1$ determined in Ref.~\cite{3loop}. From the dotted lines in 
Fig.~\ref{plots} we also see that the correction arising from the mixing 
contributions are numerically small, which is consistent with the parametric 
counting mentioned in the introduction.

It is also instructive
to analyse the impact of the NLL two-loop ultrasoft contributions in ${\cal
  V}_2^{(s)}$ and ${\cal V}_r^{(s)}$ on the NNLL evolution of the
${}^3S_1$ current coefficient $c_1$. Evaluating Eq.~(\ref{deltac1}) 
we find $(\delta c_1)^{\rm NNLL,mix}_{2,r}\approx \xi^{\rm
  NNLL}_{m;2r,{\rm usoft}}=(-1.9\%,-0.5\%)$ for 
$\nu=0.1,0.2$. These corrections lead to a partial compensation of the
anomalously large ultrasoft corrections that dominate the NNLL
non-mixing contributions to the evolution of $c_1$. It remains to be seen 
whether the two-loop ultrasoft corrections to the evolution of the
$1/(m|\bmk|)$ QCD potentials will lead to a similar effect. We will
address this issue in a subsequent work. 

\section{Conclusion}
\label{sectionconclusion}

We have determined the two-loop ultrasoft contributions to the NLL
order renormalization group equations of the ${\cal O}(v^2)$
spin-independent QCD potentials ${\cal V}_2$ ($1/m^2$) and ${\cal
  V}_r$ ($(\bmp^2+{\bmp^\prime}^2)/(2m^2 \bmk^2)$). We find
that the two-loop ultrasoft corrections are larger than the
ultrasoft contributions at the one-loop level and lead to a partial
cancellation of the anomalously large NNLL ultrasoft non-mixing
contributions to the renormalization group evolution of the 
leading ${}^3S_1$ current that describes top-antitop pair production
at threshold in $e^+e^-$ annihilation.

\acknowledgments{
This work was supported in part by the EU network contract
MRTN-CT-2006-035482 (FLAVIAnet).
}
 
\mbox{}
\vskip 1cm

\appendix

%


\begin{thebibliography}{}

\bibitem{thresholdscan}
  J.~A.~Aguilar-Saavedra {\it et al.}  [ECFA/DESY LC Physics Working Group],
  arXiv:hep-ph/0106315;
  T.~Abe {\it et al.}  [American Linear Collider Working Group],
  in {\it Proc. of the APS/DPF/DPB Summer Study on the Future of
  Particle Physics (Snowmass 2001) } ed. N.~Graf, 
  arXiv:hep-ex/0106057;
  A.~Juste {\it et al.},
  arXiv:hep-ph/0601112.


\bibitem{BBL}
  G.~T.~Bodwin, E.~Braaten and G.~P.~Lepage,
  Phys.\ Rev.\ D {\bf 51}, 1125 (1995)
  [Erratum-ibid.\ D {\bf 55}, 5853 (1997)]
  [arXiv:hep-ph/9407339].

\bibitem{synopsis}
  A.~H.~Hoang {\it et al.},
  Eur.\ Phys.\ J.\ directC {\bf 2}, 1 (2000)
  [arXiv:hep-ph/0001286].

\bibitem{HMST}
  A.~H.~Hoang, A.~V.~Manohar, I.~W.~Stewart and T.~Teubner,
  Phys.\ Rev.\ Lett.\  {\bf 86}, 1951 (2001)
  [arXiv:hep-ph/0011254];
    A.~H.~Hoang, A.~V.~Manohar, I.~W.~Stewart and T.~Teubner,
  Phys.\ Rev.\ D {\bf 65}, 014014 (2002)
  [arXiv:hep-ph/0107144].

\bibitem{PinedaSigner}
  A.~Pineda and A.~Signer,
  arXiv:hep-ph/0607239.

\bibitem{LMR} M.~Luke, A.~Manohar and I.~Rothstein,
Phys.\ Rev.\  {\bf D61}, 074025 (2000)
[arXiv:hep-ph/9910209].

\bibitem{HoangStewartultra}
A.~H.~Hoang and I.~W.~Stewart,
Phys.\ Rev.\ D {\bf 67}, 114020 (2003)
[arXiv:hep-ph/0209340].

\bibitem{amis} A.V.~Manohar and I.W.~Stewart,
Phys.\ Rev.\ D {\bf 62}, 014033 (2000)
[arXiv:hep-ph/9912226].

\bibitem{amis2}
  A.~V.~Manohar and I.~W.~Stewart,
  Phys.\ Rev.\ D {\bf 63}, 054004 (2001)
  [arXiv:hep-ph/0003107].

\bibitem{Pineda:2001et}
  A.~Pineda,
  Phys.\ Rev.\ D {\bf 66}, 054022 (2002)
  [arXiv:hep-ph/0110216].

\bibitem{HoangRuiz2}
  A.~H.~Hoang and P.~Ruiz-Femenia,
  Phys.\ Rev.\ D {\bf 74}, 114016 (2006)
  [arXiv:hep-ph/0609151].

\bibitem{3loop}
 A.~H.~Hoang,
  Phys.\ Rev.\ D {\bf 69}, 034009 (2004)
  [arXiv:hep-ph/0307376].

\bibitem{Pineda:2001ra}
  A.~Pineda,
  Phys.\ Rev.\ D {\bf 65}, 074007 (2002)
  [arXiv:hep-ph/0109117];
  N.~Brambilla, A.~Pineda, J.~Soto and A.~Vairo,
  Phys.\ Rev.\ D {\bf 60}, 091502 (1999)
  [arXiv:hep-ph/9903355].

\bibitem{Penin:2004xi}
  A.~A.~Penin, A.~Pineda, V.~A.~Smirnov and M.~Steinhauser,
  Phys.\ Lett.\ B {\bf 593}, 124 (2004)
  [arXiv:hep-ph/0403080].

\bibitem{HoangEpiphany}
  A.~H.~Hoang,
  Acta Phys.\ Polon.\ B {\bf 34}, 4491 (2003)
  [arXiv:hep-ph/0310301].

\bibitem{Penin:2004ay}
  A.~A.~Penin, A.~Pineda, V.~A.~Smirnov and M.~Steinhauser,
  Nucl.\ Phys.\ B {\bf 699}, 183 (2004)
  [arXiv:hep-ph/0406175].

\bibitem{HoangRuiz1}
 A.~H.~Hoang and P.~Ruiz-Femenia,
  Phys.\ Rev.\ D {\bf 73}, 014015 (2006)
  [arXiv:hep-ph/0511102].

\bibitem{zerobin}
  A.~V.~Manohar and I.~W.~Stewart,
  arXiv:hep-ph/0605001.

\bibitem{Grozin:2000cm}
  A.~G.~Grozin,
  arXiv:hep-ph/0008300.

\bibitem{Manohar:1997qy}
  A.~V.~Manohar,
  Phys.\ Rev.\ D {\bf 56}, 230 (1997)
  [arXiv:hep-ph/9701294].

\bibitem{StahlhofenTUM}
 M.~Stahlhofen, Diploma Thesis, Technical University Munich, 2005.

\end{thebibliography}
\end{document}